\documentclass[runningheads]{llncs}
\usepackage[T1]{fontenc}
\usepackage{graphicx}
\usepackage[hyphens]{url}
\usepackage{hyperref}
\usepackage{booktabs}
\usepackage{makecell}
\usepackage{tikz}

\usepackage{tabularx,makecell,array,ragged2e}
\newcolumntype{L}{>{\RaggedRight\arraybackslash}X}

\renewcommand\arraystretch{1.15} 

\usepackage[capitalise]{cleveref}

\makeatletter
\renewcommand{\section}{\@startsection
  {section}{1}{0mm}
  {1.5ex plus .2ex minus .2ex} 
  {0.8ex plus .1ex}            
  {\normalfont\Large\bfseries}
}
\renewcommand{\subsection}{\@startsection
  {subsection}{2}{0mm}
  {1.2ex plus .2ex minus .2ex} 
  {0.6ex plus .1ex}            
  {\normalfont\large\bfseries}
}
\makeatother

\begin{document}
\title{Early Explorations of Recommender Systems for Physical Activity and Well-being}
\titlerunning{Recommender Systems for Physical Activity}
%
\author{Alan Said\inst{1}\orcidID{0000-0002-2929-0529}}
\authorrunning{A. Said}
%
\institute{University of Gothenburg, Sweden
\email{alansaid@acm.org}
}
\maketitle              
\begin{abstract}
As recommender systems increasingly guide physical actions, often through wearables and coaching tools, new challenges arise around how users interpret, trust, and respond to this advice. This paper introduces a conceptual framework for tangible recommendations that influence users’ bodies, routines, and well-being. We describe three design dimensions: trust and interpretation, intent alignment, and consequence awareness. These highlight key limitations in applying conventional recommender logic to embodied settings. Through examples and design reflections, we outline how future systems can support long-term well-being, behavioral alignment, and socially responsible personalization.
\end{abstract}
\section{Introduction and Background}
Recommender systems have long shaped how users engage with digital media, products, and information. They are now embedded in wearables, health apps, and coaching tools that influence how we move, rest, and manage our bodies. In these settings, recommendations are not just digital suggestions. They prompt physical actions, such as what to eat, when to train, or how far to run. We refer to these as \textit{tangible recommendations}, where the consequences are embodied, cumulative, and often invisible to the algorithms that produce them.

We use ``tangible'' not in the sense of physically touchable objects, but to 
denote recommendations whose consequences manifest in physical, 
emotional, and social ways in users' embodied experience. The term distinguishes 
these from purely informational recommendations (articles, products, media) where 
consequences remain primarily cognitive or economic. We chose ``tangible'' to emphasize the concrete, felt reality of these recommendations' 
effects: sore muscles, elevated heart rate, fatigue, or injury are tangible in 
ways that reading a recommended article is not.

Previous work on health-aware or behaviorally informed recommender systems \cite{elsweiler_engendering_2016,elsweilerExploitingFoodChoice2017,schaferHealthAwareRecommender2017} has examined physical well-being, but typically treats activity as an optimization problem. This paper extends that line of research by proposing a conceptual framework for \emph{tangible recommendations} that focuses on how users interpret, trust, and embody system guidance in everyday routines.

As these systems guide offline behavior, new questions arise. What does it mean to trust a system that recommends physical activity? How can intent be inferred or adjusted when user goals shift with context? How should systems account for the longer-term consequences of their suggestions? These questions are especially relevant in domains like health, sport, and well-being, where motivation, capacity, and priorities shift frequently. A system that nudges exertion may improve fitness in one case and cause harm in another.

Although recent research has begun addressing fairness, explainability, and social impact \cite{jannach_recommender_2024}, most recommender systems still optimize for engagement rather than alignment with bodily needs. We lack conceptual tools for understanding how users interpret and act on system advice in embodied contexts. This paper develops such a tool by exploring how recommendations affect physical experience and by outlining design implications for systems that support autonomy, context awareness, and long-term well-being.

\subsection{Theoretical Positioning}
The proposed framework builds on three theoretical perspectives. From self-determination theory \cite{ryan_self-determination_2000}, it draws attention to autonomy and intrinsic motivation in physical decision-making. Research on human–automation interaction \cite{klasnja_healthcare_2012} informs the treatment of trust calibration when systems guide bodily action. Medical ethics \cite{beauchamp_principles_2019} contribute the principle of non-maleficence, relevant when algorithmic guidance entails physical risk. Together these perspectives highlight three central concerns for embodied recommendation: trust, intent, and consequence. Trust addresses the gap between system transparency and bodily awareness. Intent captures the fluidity of user goals observed in quantified-self research \cite{rooksby_personal_2014}. Consequence extends evaluation practices \cite{bauerExploringLandscapeRecommender2024,belloginImprovingAccountabilityRecommender2021} to include long-term physical and emotional outcomes. These dimensions provide the foundation for designing systems that guide physical action while respecting user autonomy.

\section{Framework: Dimensions of Tangible Recommendation}

Traditional recommender systems measure utility through relevance, engagement, or predictive accuracy. When systems begin to influence physical behavior, such as when and how users exercise, these metrics fall short. 
Tangible recommendations raise distinct design concerns. Building on prior work in health-aware recommender systems, human-computer interaction, and AI ethics, we propose a framework comprising three dimensions that define the design space of tangible recommendation: \textit{trust and interpretation}, \textit{intent and alignment}, and \textit{consequence awareness}. These relationships are illustrated in \cref{fig:tangible_dimensions}. Each dimension surfaces key tensions in designing responsible systems for physical activity and well-being, as outlined below and shown in \cref{tab:dimensions}.

\begin{figure}[t]
  \centering
  \resizebox{0.65\linewidth}{!}{
  \begin{tikzpicture}[line width=0.8pt, font=\sffamily]
    \coordinate (A) at (0,3);
    \coordinate (B) at (-2,-0.1);
    \coordinate (C) at (2,-0.1);

    \draw (A) -- (B) -- (C) -- cycle;

    \node[align=center, above] at (A) {Trust and Interpretation};
    \node[align=center, below left] at (B) {Intent Alignment};
    \node[align=center, below right] at (C) {Consequence Awareness};
  \end{tikzpicture}
  }
  \caption{Dimensions of Tangible Recommendation. Edges represent critical tensions: trust-intent (trusting intent inference), intent-consequence (goals affecting outcomes), consequence-trust (harms eroding trust).}
  \label{fig:tangible_dimensions}
\end{figure}

\subsection{Trust and Interpretation}

Trust is essential for users to act on system suggestions, especially when those suggestions affect the body. Trust is not static; it evolves through interaction, interpretation, and experience~\cite{leeTrustAutomationDesigning2004}. In tangible settings, trust depends on how users perceive system intent, how well the recommendation aligns with their state or knowledge, and how transparently it is framed.

Work in explainable AI and recommender systems shows that explanation style strongly influences trust and compliance~\cite{kunkel_let_2019,silvaLeveragingChatGPTAutomated2024,tintarev_explaining_2015}. Designing for interpretive trust involves communicating uncertainty, providing rationale, and acknowledging limits. A confident but inappropriate recommendation, e.g., such as prompting exertion during fatigue, can damage trust more than a cautious, qualified one. Presenting suggestions as guidance rather than prescriptions supports longer-term engagement~\cite{sahu_decoding_2024}.

In practice, interpretive trust can be supported through interaction design choices that make uncertainty visible and promote reflection. For example, systems may display confidence indicators or explain the limits of their models. Adaptive phrasing that changes based on context, such as distinguishing between strong and tentative recommendations, can prevent overreliance. These mechanisms help maintain user confidence while avoiding blind trust in system authority.

\subsection{Intent and Alignment}

Recommendations presume an underlying user goal, such as improving health, gaining strength, or recovering from injury. In embodied settings, intent is often fluid, context sensitive, and difficult to infer. People regularly shift between goals based on mood, fatigue, or external events. Misalignment between inferred and actual intent can lead to frustration or harm~\cite{jamesonHumanDecisionMaking2015}.

To address this, systems must support changing intent. This could involve explicit tagging (e.g., ``today I want to recover''), passive signals (e.g., reduced heart rate variability), or hybrid models. Without such flexibility, recommendations risk becoming irrelevant or counterproductive.

Intent alignment can be operationalized through a combination of explicit and implicit inputs. Users might signal their daily goals through simple modes such as ``train,'' ``recover,'' or ``rest,'' while the system interprets physiological data or contextual cues to adjust its suggestions. This combination allows personalization to evolve as goals shift, keeping recommendations appropriate to the user’s physical and emotional state.

\subsection{Consequence Awareness}

Tangible recommendations have effects that go beyond engagement. Physical outcomes (e.g., injury), emotional responses (e.g., guilt), and social dynamics (e.g., peer pressure) are rarely captured in standard evaluation metrics~\cite{bauerExploringLandscapeRecommender2024}.

Sustainable systems must be sensitive to downstream effects, even if they cannot predict them exactly. Prior work in health-aware recommender systems emphasizes the need to balance engagement with long-term well-being~\cite{schaferHealthAwareRecommender2017}, and recent evaluation research highlights the importance of accounting for context-specific harms~\cite{belloginImprovingAccountabilityRecommender2021}. This could include support for recovery, more varied recommendations, and mechanisms that encourage reflection~\cite{bauerExploringLandscapeRecommender2024,schaferHealthAwareRecommender2017,belloginImprovingAccountabilityRecommender2021}.

Designing for consequence awareness means accounting for the outcomes of repeated system influence, not only immediate engagement. Interfaces can invite users to provide feedback about how activities felt or whether they matched expectations. Over time, such reflective data could help identify unintended effects such as fatigue, guilt, or overtraining. Including mechanisms for users to adjust or pause recommendations contributes to safer and more sustainable system behavior.

\begin{table}[ht]
\centering
\caption{Design dimensions for tangible recommendations}
\label{tab:dimensions}
\begin{tabular}{p{0.25\linewidth} p{0.67\linewidth}}
\toprule
\textbf{Dimension} & \textbf{Design Consideration} \\
\midrule
\makecell[tl]{\textbf{Trust and}\\ \textbf{Interpretation}} &
Support transparency, calibrate confidence, communicate uncertainty; frame recommendations as guidance, not prescription. \\
\hline
\makecell[tl]{\textbf{Intent and}\\ \textbf{Alignment}} &
Enable dynamic goal-setting and intent signaling; design for fluid, situational motivation rather than static profiles. \\
\hline
\makecell[tl]{\textbf{Consequence}\\ \textbf{Awareness}} &
Consider physical and emotional impact; support recovery, routine variation, and reflective engagement with long-term outcomes. \\
\bottomrule
\end{tabular}
\end{table}

\subsection{Design Patterns}
To operationalize the framework, we outline design patterns that illustrate how trust, intent, and consequence can be addressed in practice.

\paragraph{\textbf{Trust and Interpretation}}
Confidence displays can express the reliability of data and inferences, for example, ``Based on your HRV (confidence: moderate), intervals might work today.'' Easy override options allow users to reject suggestions without penalty, and guidance framing encourages autonomy by phrasing recommendations as possibilities rather than prescriptions.

\paragraph{\textbf{Intent Alignment}}
Systems can support simple daily goal modes (e.g., Train, Maintain, Recover) that reframe all subsequent recommendations. Contextual inference, such as suggesting recovery after poor sleep or stress, helps align with situational needs. Multi-goal modeling allows users to balance overlapping objectives, for instance combining endurance and flexibility targets.

\paragraph{\textbf{Consequence Awareness}}
Designs can provide post-activity reflections that capture affective and physical feedback (How did this feel?), visualize cumulative load with recovery indicators, and frame rest as part of sustainable progress rather than interruption.\\

These patterns suggest corresponding metrics: for trust, override frequency and calibration accuracy; for intent, goal-recommendation alignment and user-reported relevance; and for consequence, physical and emotional well-being outcomes such as injury rates, long-term retention, or perceived energy balance.

\section{Illustrative Scenarios}

To ground the framework, we present three scenarios that illustrate how tangible recommendations relate to trust, intent, and consequence in everyday physical activity, as summarized later in \cref{tab:scenario_mapping}. While hypothetical, they reflect patterns from real-world use.

\subsection{Scenarios}

\subsubsection*{Scenario 1: Overriding the Body}

Maria, a recreational runner, uses a smartwatch with a built-in virtual coach. After a week of intense training, the system recommends a high-intensity interval workout based on her heart rate variability data. Although Maria feels physically exhausted, she interprets the recommendation as a sign she \emph{should} push harder. She completes the session but suffers muscle strain that prevents her from running for two weeks.

This scenario highlights the risks of over-trusting system authority and the absence of bodily context in the recommendation logic. It exemplifies failures in both \emph{interpretive trust} and \emph{consequence awareness}~\cite{leeTrustAutomationDesigning2004,schaferHealthAwareRecommender2017}.

\subsubsection*{Scenario 2: Misaligned Recovery Intent}

Carl is recovering from a knee injury. His wearable has learned his preference for long-distance cycling and continues to recommend hilly rides. However, his physiotherapist advises low-resistance indoor training. Without a way to express this change in intent, Carl disregards the system altogether.

This case shows how static user modeling fails with shifting needs. The lack of support for \emph{goal fluidity} and \emph{intent realignment} leads to disuse and distrust~\cite{jamesonHumanDecisionMaking2015}.

\subsubsection*{Scenario 3: Invisible Consequences in Group Training}
Harper participates in a team-based training program where everyone’s wearable data feeds into a shared dashboard. The system recommends pace adjustments during runs to optimize group performance. However, Harper, who is less fit than others, follows the recommendation and feels ashamed when not being able to keep up and later skips a session without telling the group.

This case highlights the social and emotional consequences of seemingly neutral system logic. Recommendation quality cannot be assessed solely by performance metrics---the \emph{social dynamics and psychological load} of recommendations must also be considered~\cite{sansonettiDynamicSocialRecommendation2017,quijano-sanchezSocialFactorsGroup2013}.

\subsection{Design Insights}
These scenarios show how tangible recommendations challenge assumptions about personalization and system authority. In Maria’s case, trust becomes problematic not due to system failure, but because persuasive recommendations override her bodily awareness. The system is technically correct but contextually misaligned. In Carl’s case, the system misses a shift in intent, leading to irrelevant advice and eventual disengagement. Harper’s experience adds another layer, revealing how social visibility and group dynamics can amplify the emotional consequences of system guidance. Together, these examples point to the need for calibrated trust, adaptable intent modeling, and attention to social as well as physical context.

Intent must be treated as fluid, since physical condition, motivation, and external constraints shift in ways that static personalization cannot capture. Without mechanisms for users to revise or suppress inferred goals, recommendations may become counterproductive. The consequences of recommendation are therefore not only behavioral but also emotional and social: when systems ignore recovery needs, reinforce pressure, or limit autonomy, they may erode well-being rather than support it.

Tangible recommender systems should therefore be designed with contextual awareness and sensitivity to consequence. Supporting well-being in physical activity involves more than matching preferences; it requires sustained alignment with users’ changing capacities, goals, and interpretations of movement.

\cref{tab:scenario_mapping} summarizes how the three scenarios reflect the framework’s three dimensions. Maria’s case (Scenario 1) highlights a breakdown in interpretive trust and a lack of consequence awareness, where the system’s persuasive framing overrides bodily context. Carl’s case (Scenario 2) illustrates a failure to capture fluid intent, showing how static personalization can lead to disengagement. Harper’s case (Scenario 3) highlights how group-oriented recommendations introduce social and emotional consequences that extend beyond individual control. Mapping scenarios to framework dimensions clarifies how specific design choices affect trust, alignment, and long-term outcomes.

\begin{table}[t]
\centering
\caption{Mapping of framework dimensions to the three illustrative scenarios.}
\setlength{\tabcolsep}{4pt}
\renewcommand{\arraystretch}{1.15}
\resizebox{\linewidth}{!}{
\begin{tabular}{@{} l l l l @{}}
\toprule
\textbf{Dimension} &
\textbf{\shortstack[tc]{Scenario 1}} &
\textbf{\shortstack[tc]{Scenario 2}} &
\textbf{\shortstack[tc]{Scenario 3}} \\
\midrule
\makecell[tl]{\textbf{Trust and}\\ \textbf{Interpretation}} &
\makecell[tl]{Overreliance on system \\authority; limited com-\\munication of uncertainty.} &
\makecell[tl]{Erosion of trust after\\ repeated misfit\\ recommendations.} &
\makecell[tl]{Trust shaped by peer\\ visibility and shared\\ performance data.} \\
\hline
\makecell[tl]{\textbf{Intent}\\ \textbf{Alignment}} &
\makecell[tl]{Goal partially inferred\\ but insensitive to fatigue\\ or recovery state.} &
\makecell[tl]{No mechanism for\\ expressing changing\\ goals or recovery intent.} &
\makecell[tl]{Group optimization\\ overrides individual\\ motivation or capacity.} \\
\hline
\makecell[tl]{\textbf{Consequence}\\ \textbf{Awareness}} &
\makecell[tl]{Neglects risk of strain\\ and reduced motivation.} &
\makecell[tl]{Long-term disengage-\\ment due to unacknow-\\ledged recovery needs.} &
\makecell[tl]{Social pressure and\\ emotional discomfort\\ lead to avoidance\\ or dropout.} \\
\bottomrule
\end{tabular}
}
\label{tab:scenario_mapping}
\end{table}

\section{Positioning and Outlook}
The scenarios discussed above demonstrate how recommender systems in embodied contexts raise new challenges for personalization. Although wearables increasingly offer activity suggestions, we still know little about how different types of users interpret and act on this guidance. Athletes, patients, and casual exercisers may have very different relationships to system authority, bodily feedback, and recovery needs \cite{rapp_personal_2018}. Addressing this gap will require empirical studies that attend to lived experience through interviews, diary methods, and reflective data collection. System design must also enable users to shift, suspend, or redefine their goals as physical and emotional states change.

Although this paper focuses on physical activity, the same framework applies to other embodied domains where recommendations influence real-world behavior, such as food choice, rehabilitation, sleep, or mental well-being. These settings share similar tensions around trust, intent, and consequence, as systems interpret user states and guide action over time. Extending the framework to these contexts can help reveal domain-specific challenges while supporting a consistent approach to responsible, human-centered personalization.

Evaluating these systems will require a rethinking of priorities. Metrics such as click-through rate or precision offer little insight into the value or risk of a recommendation that affects the body. Future work should explore evaluation approaches that reflect long-term well-being, behavioral alignment, and user agency. These include consequence-aware evaluation techniques, subjective trust assessments, and transparency methods that help users understand what is being recommended and why.

This framework contributes to a broader research agenda in sustainable and responsible recommendation. In the domain of physical activity, sustainability refers not to computational efficiency, but to the preservation of health, motivation, and trust over time. Designing for these outcomes means attending to users' fluctuating goals, embodied constraints, and personal meanings of movement. The next step is empirical work that examines how people interpret and adapt to tangible recommendations in daily life and how evaluation can capture both immediate utility and lasting well-being. Such research can guide the development of recommender systems that remain effective while respecting user autonomy and lived experience.

%
%
%
\bibliographystyle{splncs04}
\bibliography{bibliography}
\end{document}